# Heterointerface Control over Lithium-induced Phase Transitions in MoS$_2$ Heterostructures

Joshua V. Pondick[1,2], Aakash Kumar[1,2], Mengjing Wang[1,2], Sajad Yazdani[1,2], John M. Woods[1,2], Diana Y. Qiu[1,2], Judy J. Cha[1,2]*

[1]Department of Mechanical Engineering and Materials Science, Yale University, New Haven, CT 06511, USA.

[2]Energy Sciences Institute, Yale West Campus, West Haven, CT 06516, USA.

*Correspondence to: judy.cha@yale.edu

**Abstract**

Phase transitions of two-dimensional materials and their heterostructures enable many applications including electrochemical energy storage, catalysis, and memory; however, the nucleation pathways by which these transitions proceed remain underexplored, prohibiting engineering control for these applications. Here, we demonstrate that the lithium intercalation-induced 2H-1T' phase transition in $MoS_2$ proceeds via nucleation of the 1T' phase at a heterointerface by monitoring the phase transition of $MoS_2$/graphene and $MoS_2$/hexagonal boron nitride (*h*BN) heterostructures with Raman spectroscopy *in situ* during intercalation. We observe that graphene-$MoS_2$ heterointerfaces require an increase of 0.8 V in applied electrochemical potential to nucleate the 1T' phase in $MoS_2$ compared to *h*BN-$MoS_2$ heterointerfaces. The increased nucleation barrier at graphene-$MoS_2$ heterointerfaces is due to the reduced charge transfer from lithium to $MoS_2$ at the heterointerface as lithium also dopes graphene based on *ab initio* calculations. Further, we show that the growth of the 1T' domain propagates along the heterointerface, rather than through the interior of $MoS_2$. Our results provide the first experimental observations of the heterogeneous nucleation and growth of intercalation-induced phase transitions in two-dimensional materials and heterointerface effects on their phase transition.

**Main Text**

Nucleation phenomena are fundamental to many natural processes, and remain a critical area of study in a wide range of fields including protein phase transitions in biology,[1–4] nucleation of atmospheric particulates in climate science,[5,6] crystallization dynamics in metals for structural materials,[7–9] and synthesis of nanomaterials.[10–12] Phase transitions are initiated by nucleation, and the transition pathways between phases are complex and heavily influenced by factors such as nanoscale confinement and interfaces.[13] For layered transition metal dichalcogenides (TMDs) that can exist in several structural polymorphs,[14] the phase transitions between these polymorphs have proven important for neuromorphic memristive computing,[15–17] catalytic production of hydrogen,[18–22] and logic devices.[23] To effectively exploit these phase transitions for these applications, understanding the thermodynamics and kinetics of the nucleation pathways is essential; however, nucleation pathways remain virtually unexplored for many TMDs that have been demonstrated to undergo phase transitions.

For example, despite the well-established phase transition from the semiconducting 2H phase to the semimetallic 1T' phase via lithium intercalation into the TMD $MoS_2$,[24–27] the nucleation pathway has not been experimentally studied at the microscopic level. *Ab initio* calculations indicate that the lithium-induced phase transition in $MoS_2$ is due to the electron doping from intercalated lithium atoms into the conduction band of 2H-$MoS_2$: at a critical donated electron concentration, the sulfur atoms change their coordination around molybdenum from trigonal prismatic to octahedral, thus forming 1T'-$MoS_2$.[28–31] Constructing heterointerfaces by interfacing $MoS_2$ with other two dimensional (2D) materials could modulate the charge donation from lithium at the heterointerface, thus enabling the study of nucleation pathways in the lithium-induced 2H-1T' phase transition of $MoS_2$. Recently, heterointerface effects in the electrochemical intercalation of van der Waals (vdW) heterostructures have been investigated to understand how heterointerfaces modulate electron transport and lithium storage.[32] However, heterointerface effects on the phase transition, and consequently on the nucleation pathway, have not been examined.

We investigate the effects of a heterointerface on the phase transition of $MoS_2$ by considering the lithium intercalation of two heterostructures: graphene-$MoS_2$ and hexagonal boron nitride ($h$BN)-$MoS_2$. Upon lithium intercalation, graphene readily accepts charge from lithium,[33,34] while insulating $h$BN is not expected to interact significantly with intercalated lithium.[35,36] We employed density functional theory (DFT) calculations to determine the equilibrium configuration of the graphene-$MoS_2$ and $h$BN-$MoS_2$ heterointerfaces with a Li intercalant, followed by Bader charge analysis to determine how charge from the Li is donated at each heterointerface. As shown in Figure 1, Li donates close to 1 electron to freestanding $MoS_2$. Consistent with our previous work,[36] the amount of charge donated by Li is slightly reduced at the $h$BN-$MoS_2$ heterointerface, with 90% of the charge donated to the $MoS_2$ layer and 10% donated to the $h$BN. At the graphene-$MoS_2$ heterointerface, however, we find a dramatic reduction in charge donation from Li to $MoS_2$, with 34% of the charge transferred to the $MoS_2$ layer and 48% to the three adjacent graphene layers (the rest of the charge (18%) is donated to the bottom three graphene layers (now shown in the schematic in Fig. 1b)). Density of state calculations show that lithium can dope electrons across a wide range of energies at the $MoS_2$-graphene heterointerface (Supplementary Fig. 1). Thus, we find that interfacing $MoS_2$ with graphene lowers the overall charge donated, and hence the effective doping power of Li, to $MoS_2$ by ~60% when compared to both the $MoS_2$-$h$BN heterostructure and freestanding $MoS_2$.

Motivated by the DFT analysis, we fabricated heterostructures of several-layer $MoS_2$ interfaced with graphene and $h$BN using mechanically exfoliated microflakes patterned with Cr/Au electrical contacts and supported on $SiO_2$/Si substrates. These devices were integrated into electrochemical microreactors,[32,34,36–38] in which the Cr/Au-contacted heterostructures served as the working electrode, while lithium metal pressed onto copper foil served as the counter electrode (Supplementary Fig. 2). These electrodes were immersed in a liquid electrolyte (LiPF$_6$ in EC/DEC) and sealed in an airtight cell fitted with a glass window to allow for *in situ* optical and Raman characterization during intercalation. Intercalation was controlled potentiostatically by sweeping the electrochemical potential (V$_{EC}$) between the working electrode and lithium metal. As V$_{EC}$ was sequentially lowered from open circuit voltage (OCV, ~2.5 V vs. Li/Li$^+$) to 0.2 V vs. Li/Li$^+$ at a decrement of 0.2 V, Li$^+$ ions diffused through the electrolyte and intercalated into the interlayer gaps of the heterostructures. At each potential, V$_{EC}$ was held constant to acquire Raman spectra and optical micrographs *in situ*.

To investigate the effect of a heterointerface between MoS$_2$ and graphene, we partially covered a MoS$_2$ flake with graphene to create bare and graphene-covered regions of MoS$_2$ (Fig. 2a,b). V$_{EC}$ was lowered from OCV to 1.0 V vs. Li/Li$^+$, which is the potential at which the 2H-1T' phase transition is expected to occur for bare MoS$_2$.[36–38] *In situ* optical microscopy revealed that the uncovered portion of MoS$_2$ began to darken at 1.0 V vs. Li/Li$^+$ (Fig. 2c), indicative of the 2H-1T' phase transition.[36,38] After 120 minutes at 1.0 V vs. Li/Li$^+$, the bare region of MoS$_2$ completely darkened. Strikingly, the graphene-covered region retained its original blue color even after 120 minutes at 1.0 V vs. Li/Li$^+$ (Fig. 2c), suggesting that the graphene-covered MoS$_2$ remained in the 2H phase. *In situ* Raman spectra confirm that the uncovered region of the MoS$_2$ flake underwent a phase transition at 1.0 V vs. Li/Li$^+$ as evidenced by the disappearance of the E$_{2g}$ and A$_{1g}$ modes of 2H-MoS$_2$ and the concurrent growth of the J$_1$ and J$_2$ modes of 1T'-MoS$_2$ (Fig. 2d).[23,39] By contrast, *in situ* Raman spectra confirm that the graphene-covered region remained in the 2H phase after 120 minutes at 1.0 V vs. Li/Li$^+$ (Fig. 2e). As V$_{EC}$ was further lowered to 0.9 V, 0.8 V, and 0.6 V vs. Li/Li$^+$, the darkened region expanded inwards from the bare region to the graphene-covered region of MoS$_2$. However, the growth of the darkened region was slow, and after 20 minutes at 0.6

V vs. Li/Li+, only half of the graphene-covered region had undergone the phase transition. At 0.4 V vs. Li/Li+, the graphene-covered $MoS_2$ completed the phase transition to 1T', while the bare region of $MoS_2$ was fully converted to Mo and $Li_2S$.[37,40] For the graphene-covered region, the irreversible conversion reaction occurred at 0.2 V vs. Li/Li+. Thus, Figure 2 shows that the graphene-$MoS_2$ heterointerface significantly delayed the phase transition.

Analysis of *in situ* optical micrographs elucidates the heterointerface effects on the growth dynamics of the 1T' domain (Supplementary Fig. 3). At 1.0 V vs. Li/Li+, the bare region of $MoS_2$ darkened quickly inwards from the flake edges in contact with the electrolyte, suggesting nucleation of 1T' domains at the edges, followed by fast growth of the 1T' domains (Fig. 2f-i and 2f-ii). At 0.6 V vs. Li/Li+, the growth of the dark region in the graphene-covered portion of the flake continues slowly from the uncovered region. Thus, when 1.0 V > $V_{EC}$ > 0.4 V vs. Li/Li+, 1T' domains do not nucleate in the graphene-covered region, while growth of already-nucleated 1T' domains is extremely slow (Fig. 2f-iii). At 0.4 V vs. Li/Li+, the growth of the 1T' domain accelerated significantly and also proceeded from the graphene-covered edges inwards (Fig. 2f-iv), until the entire covered-region of $MoS_2$ fully converted to the 1T' phase (Supplementary Fig. 3). The graphene-$MoS_2$ heterointerface effects on the phase transition of $MoS_2$ were reproduced in a replicate device (Supplementary Fig. 4). We note that *in situ* Raman spectroscopy is sufficient to study these phase transitions as Raman spectra for the 2H and 1T' phases are distinct. *In situ* x-ray diffraction would produce too little signal for the microflakes under study, while *in situ* transmission electron microscopy is challenging with liquid electrochemistry.

From these experiments, we conclude that the 2H-1T' phase transition initiates via heterogenous nucleation. Our *ab initio* calculations show the electron doping power of lithium to $MoS_2$ is reduced at the graphene-$MoS_2$ heterointerface, suggesting that a greater applied $V_{EC}$, and thus a higher lithium concentration, is required to achieve the same donated electron concentration at the graphene-$MoS_2$ heterointerface. However, only the top-most $MoS_2$ layer is expected to lose electron density to graphene, leaving the doping power of lithium in the interior vdW gaps of $MoS_2$ unchanged. Since our flakes are thicker than 5 layers, the heterointerfaced layer represents less than 20 % of the volume. If the nucleation of 1T' phase were homogeneous, i.e., from the interior volume of $MoS_2$, we would expect a mixed Raman spectrum of both the 2H and 1T' phases; however, we do not observe the coexistence of both phases. Therefore, we conclude that the phase transition must initiate at the heterointerface between graphene and $MoS_2$. Furthermore, the growth of the 1T' phase must also be heterogenous, propagating along the heterointerface, rather than through the interior of the flake. At 1.0 V vs. Li/Li$^+$, the graphene-covered region of $MoS_2$ remained in the 2H-phase while the uncovered region fully converted to the 1T' phase, forming a lateral 2H-1T' heterointerface (Fig. 2f-ii). If growth of the 1T' phase progressed homogenously, then the 1T' phase would be expected to grow from the interior layers at 1.0 V vs. Li/Li$^+$ regardless of the heterointerface at the uppermost layer, which we do not observe. This observation is consistent with previous electron diffraction analysis of intercalated bulk $MoS_2$ using *in situ* TEM, which suggested that the 2H-1T' phase transition front propagated along the top and bottom free surfaces before the interior regions.[41]

The nucleation barrier for the 1T' phase can thus be modulated by a heterointerface. To further investigate the heterogeneous nucleation, we fabricated a graphene/$MoS_2$/graphene heterostructure in which $MoS_2$ is completely encapsulated by top and bottom graphene flakes with two heterointerfaces (Fig. 3a,b). During the intercalation of lithium into this heterostructure, *in situ* optical microscopy and Raman spectroscopy revealed that the $MoS_2$ flake remained in the 2H phase even after $V_{EC}$ was lowered to 0.4 V vs. Li/Li$^+$ (Fig. 3c,d), an applied voltage at which pristine $MoS_2$ breaks down to Mo clusters and $Li_2S$ (Fig. 2d). The 1T' phase was finally nucleated when $V_{EC}$ was lowered to 0.2 V vs. $Li/Li^+$, at which point the flake darkened (Fig. 3c) and the $J_1$ and $J_2$ peaks of 1T'-$MoS_2$ appeared (Fig. 3d). The delayed phase transition might be because the $MoS_2$ flake was not directly exposed to the liquid electrolyte. However, in our previous study, we showed that $MoS_2$ still underwent the phase transition at 1.0 V vs. $Li/Li^+$ despite not being exposed to the electrolyte.[36] Further intercalation of this heterostructure was not possible due to the destruction of the electrical contacts from the alloying of gold with lithium at 0.2 V vs. $Li/Li^+$.[42,43] Intercalation of a replicate device showed similar results (Supplementary Fig. 5; lithium intercalation into the graphene flakes is shown in Supplementary Fig 6). Thus, using two differently configured heterostructures (Fig. 2 and Fig. 3), we clearly demonstrated a delay in the phase transition of up to 0.8 V, which we attribute to the heterogeneous nucleation and heterogeneous growth of the 1T' phase.

We next investigated the growth kinetics of the 1T' phase at different heterointerfaces by intercalating a heterostructure of $MoS_2$ partially supported on graphene and partially on $SiO_2$ where the top basal plane of $MoS_2$ was exposed to the electrolyte to initiate the nucleation at the top-most layer (Fig. 4a,b). Surprisingly, the bottom graphene-$MoS_2$ heterointerface accelerated the phase transition instead of delaying it. *In situ* optical micrographs and Raman spectra shown in Fig. 4c-e reveal that the graphene-supported region of the $MoS_2$ flake not only underwent the phase transition at 1.0 V vs. $Li/Li^+$ as expected due to the free top-most layer of $MoS_2$, but did so faster than the $SiO_2$-supported region of $MoS_2$. The graphene-supported region completed the phase transition after 30 minutes at 1.0 V vs. $Li/Li^+$, while the $SiO_2$-supported region did so after 70 minutes at 1.0 V vs. $Li/Li^+$. Both regions underwent the irreversible conversion reaction as expected at 0.4 V vs. $Li/Li^+$ (Fig. 4d-e), and we observed the same behavior in a replicate device (Supplementary Fig. 7). This indicates that the mere presence of a graphene-$MoS_2$ heterointerface is not sufficient to delay the phase transition.

To understand the faster growth kinetics at a $MoS_2$-graphene heterointerface at the bottom of the $MoS_2$ flake, we considered two possibilities. The first is more efficient electron injection into $MoS_2$ for faster intercalation kinetics as graphene makes superior electrical contact to $MoS_2$ than gold does due to a reduced Schottky barrier at the interface.[44,45] Thus, it is possible that the graphene-$MoS_2$ heterointerface facilitated more efficient electron injection than the gold electrode, allowing for a more rapid phase transition to the 1T' phase. However, intercalation of a $MoS_2$ flake partially supported on graphene and biased only through $MoS_2$ showed the same enhancement in the phase transition kinetics, eliminating the possibility of enhanced charge injection (Supplementary Fig. 8). The second possibility is that the graphene support could facilitate the release of mechanical strain induced by the phase transition. The 2H-1T' phase transition induces mechanical strain, which can result in the formation of wrinkles and a buckled microstructure.[38] *Post mortem* scanning electron microscopy (SEM) of the $MoS_2$-graphene heterostructures after intercalation revealed that the formation of wrinkles and buckled microstructures was more apparent on the graphene-supported regions of $MoS_2$ as compared to the $SiO_2$-supported regions (Supplementary Fig. 9 and Fig. 10). Therefore, we conclude that the $MoS_2$-graphene heterointerface can relieve the strain caused by the phase transition more easily than the $MoS_2$-$SiO_2$ interface, leading to faster kinetics of the phase transition.

Our *ab initio* calculations suggest that while the graphene-$MoS_2$ heterointerface will reduce lithium's electron doping power to $MoS_2$, an $h$BN-$MoS_2$ interface has little effect on the doping power of lithium. To probe this, we fabricated a heterostructure where $MoS_2$ was partially covered by $h$BN (Fig. 5a,b), and observed that the $h$BN-covered $MoS_2$ underwent the phase transition at 1.0 V vs. Li/Li$^+$ (Fig. 5c,d and Supplementary Fig. 11a), in agreement with our calculations and our previous findings.[36] We also probed the kinetics of the phase transition at the heterointerface by fabricating a heterostructure where MoS$_2$ was partially supported on $h$BN and partially on SiO$_2$, with the free top basal plane of MoS$_2$ exposed to the electrolyte (Fig. 5e,f). Just as in the graphene-supported case, the $h$BN-supported MoS$_2$ underwent the phase transition more rapidly at 1.0 V vs. Li/Li$^+$ than the SiO$_2$-supported region (Fig. 5g,h and Supplementary Fig. 11b). *Post mortem* SEM of this heterostructure revealed a more pronounced wrinkled microstructure of the $h$BN-supported region as compared to the SiO$_2$-supported region (Supplementary Fig. 11c-e). Therefore, we show that interfacing MoS$_2$ with a 2D substrate allows for a more rapid phase transition due to the efficient release of mechanical strain due to weak vdW interactions at the 2D heterointerface.

In summary, our results show that the lithium-induced phase transition in MoS$_2$ proceeds by heterogeneous nucleation of the 1T' phase at a heterointerface rather than homogeneous nucleation, and the nucleation barrier for the 1T' phase is the highest for the graphene-MoS$_2$ interface, requiring an increase in V$_{EC}$ as high as 0.8 V more than for free MoS$_2$ surface. The growth kinetics of the phase transition are also influenced by the heterointerface, where 2D vdW interfaces facilitate faster growth kinetics than MoS$_2$ supported on SiO$_2$. Our results thus provide microscopic insight into the nucleation pathways for phase transitions in 2D materials and highlight the importance of heterointerfaces on the onset and propagation of phase transitions. Particularly, as graphene and $h$BN are widely employed as nanoscaled electrodes and passivation layers for many 2D heterostructures, understanding their heterointerfacial effects on phase transition dynamics of the active material is of paramount importance.

**Methods**

### *Ab Initio Calculations*

Density Functional Theory calculations were carried out using a plane-wave basis set within the Projector Augmented Wave[46,47] approach using the Quantum Espresso[48] software package. The $4s$ and $4p$ semi-core states of Mo were included as valence electrons and the exchange-correlation was treated at the Generalized Gradient Approximation (GGA) level of Perdew-Becke-Ernzehof (PBE).[49] The vdW interactions were accounted for using Grimme's D3 dispersion correction.[50] A 4x4 $MoS_2$-5x5 graphene supercell was used containing 1 monolayer of $MoS_2$ and 6 layers of graphene, where the graphene was strained in-plane by ~-1.3% to create a commensurate interface, with zero strain in the direction normal to the interface. The kinetic energy cut-off for the plane waves was set to 1040 eV, and a Gamma centered Monkhorst-Pack[51] K-mesh of 8x8x1 was employed to sample the Brillouin Zone. A Li atom was introduced at the preferred site on top of a Mo atom[52] and all the atoms in the supercell were relaxed until the total energy converged to within 0.2 meV/atom, and the forces on each atom were smaller than 0.03 eV/Å. Bader charge analysis[53] was carried on the relaxed equilibrium configuration to determine the change in the charge distribution induced by Li. For freestanding $MoS_2$, Li donated -0.88$e$ (where $e$ stands for the elementary charge), while in a graphene/$MoS_2$ heterostructure, Li donated -0.30$e$ to $MoS_2$ and -0.42$e$ to the three graphene layers adjacent to Li, and -0.16$e$ to the bottom graphene layers. For the hBN-$MoS_2$ heterointerface, a 4x4 $MoS_2$-5x5 $h$BN supercell was used containing 1 monolayer of $MoS_2$ and 6 layers of $h$BN, where the $h$BN was strained in-plane by ~-1.3% to create a commensurate interface. Further details for the $h$BN-$MoS_2$ calculations can be found in our previous work.[36] Bader charge analysis revealed that for freestanding $MoS_2$, Li donated -0.97$e$, while in a $h$BN/$MoS_2$ heterostructure, Li donated -0.87$e$ to $MoS_2$ and -0.1$e$ to $h$BN.

### *Device Fabrication*

Several-layer $MoS_2$ (SPI Supplies), graphene (NGS Naturgraphit GmbH), and $h$BN (HQ Graphene) flakes were mechanically exfoliated from bulk crystals onto $SiO_2$/Si substrates using the scotch-tape method. The substrates were sonicated in acetone and isopropyl alcohol, and treated with $O_2$ plasma prior to exfoliation. $MoS_2$ flake thickness was identified via the separation between the $E_{2g}$ and $A_{1g}$ Raman modes[54,55] measured using a Horiba LabRAM HR Evolution Spectrometer with a 532 nm laser and a 1800 lines/mm diffraction grating (Supplementary Fig. 2). Graphene flake thickness was determined via the ratio of the intensity of the Si Raman peak at 520 $cm^{-1}$ measured with a 633 nm laser through the flake as compared to the intensity of the Si peak measured directly from the substrate as we previously describe.[56] Raman analysis shows a low defect density in the graphene flakes (Supplementary Fig. 2). $h$BN flake thickness was estimated using optical microscopy.

Flakes of desired size and thickness were transferred to $SiO_2$/Si substrates using a KOH-assisted technique, as we describe previously[36,56]. Briefly, a hemispherical droplet of epoxy (Scotch-Weld, Series DP100Plus) on a glass slide was coated with a layer of polypropylene carbonate (PPC, Sigma Aldrich). Using an optical microscope and a micro-manipulator that holds the glass slide, the epoxy/PPC droplet was positioned above a flake of interest and carefully lowered to contact the flake. A 2M aqueous solution of potassium hydroxide (KOH, Sigma Aldrich) was added to the substrate to etch the top few Å of $SiO_2$, releasing the flake from the substrate onto the epoxy/PPC droplet. Each flake was released from the glass slide by melting the PPC at 95-100 °C for 5 minutes while contacting a target $SiO_2$/Si substrate. The PPC was subsequently dissolved in chloroform overnight. Heterostructures were fabricated by repeating this procedure and using the micro-manipulator and rotational stage to carefully control the placement of the second flake on top of the previously-transferred bottom flake. All $MoS_2$ flakes used in this study were 5-7 layers thick, all graphene support flakes were 6-10 layers thick, and all graphene flakes covering $MoS_2$ were 5 layers thick to minimize the attenuation of Raman signal from $MoS_2$.

For electrochemical lithium intercalation, electrodes were patterned with electron beam lithography (Nabity NPGS, Helios G4 FIB-SEM) and then 10 nm Cr / 100 nm Au was deposited using thermal evaporation (Mbraun EcoVap). For partially-covered $MoS_2$/graphene devices, electrodes were patterned onto both $MoS_2$ and graphene; for $MoS_2$ encapsulated by graphene, electrodes were deposited onto both the top and bottom graphene flakes; and for $h$BN/$MoS_2$ heterostructures, electrodes were deposited onto $MoS_2$ only.

### *Electrochemical Cell Fabrication*

Cr/Au-contacted heterostructures fabricated on $SiO_2$/Si were attached to a glass slide, and the gold contacts of the devices were wire-bonded to aluminum tape for connection to electrical instrumentation. All subsequent steps were conducted in an argon glovebox. For all experiments, intercalation was conducted with a liquid electrolyte using an enclosed cell that holds the device and electrolyte and is sealed with an optical-grade glass top cover.[36–38,57] Three sides of the glass top cover were first sealed by epoxy (Scotch-Weld, Series DP100Plus), leaving one side open. After the epoxy cured, a small piece (~3×3 mm) of lithium metal (0.38 mm-thick ribbon, Sigma-Aldrich) was pressed onto copper foil using a mechanical plier to ensure good contact. The lithium/copper foil was then inserted into the open side of the glass top cover. The liquid electrolyte, a battery-grade solution of 1 M lithium hexafluorophosphate in 50/50 v/v ethylene carbonate / diethyl carbonate (LiPF$_6$ in EC/DEC, Sigma Aldrich), was added to the cell to submerge the device and lithium metal. The open side was then sealed with epoxy, forming an air-tight seal.

### *In Situ Raman Characterization During Intercalation*

Intercalation cells were connected to a Biological SP300 potentiostat/galvanostat for the electrochemical intercalation of Li$^+$. The Cr/Au contacts to the device served as the working electrode, while the lithium/copper served as the reference/counter electrode. In all experiments, working electrodes were connected together to allow for simultaneous intercalation through all working electrodes. Before intercalation, a Raman spectrum was taken at OCV (typical OCV values were 2.6 – 2.9 V vs. Li/Li$^+$). Lithium was intercalated into the heterostructures potentiostatically by dropping $V_{IC}$ vs. Li/Li$^+$ at a scan rate of 10 mV s$^{-1}$. Upon reaching a desired $V_{IC}$, the cell was held at that potential while Raman spectra were collected. For multiple intercalation cycles, cells were allowed to recover to OCV before the next intercalation.

All Raman spectra were taken with a Horiba LabRAM HR Evolution Spectrometer using a 633 nm HeNe laser with an 1800 lines/mm diffraction grating. Before intercalation, all samples were characterized at a laser power of ~3 mW to avoid damage, but after cell fabrication, a laser power of ~7.5 mW was used to increase the signal-to-noise ratio due to scattering by the electrolyte. *In situ* Raman spectra were collected with fifteen 5-second exposures. We note that due to extremely low signal, *h*BN flakes could not be characterized at this Raman power (Supplementary Fig. 2). Raman spectroscopy was used to monitor the phase of intercalated MoS$_2$ as it is a non-invasive technique that can sample the phase throughout the entire thickness of the heterostructures with micron-scale spatial resolution. Other characterization techniques such as *in situ* transmission electron microscopy are incompatible with liquid electrolytes, while x-ray diffraction or photoelectron spectroscopy conducted on microflakes yield extremely low signals that prevent meaningful analysis.

### *Post Mortem Characterizations*

Post intercalation, the microreactors were opened using a razor blade to remove the epoxy walls to recover the $SiO_2$/Si on-chip device. Recovered devices were placed into an isopropyl alcohol wash and dried, and then characterized with optical microscopy and Raman spectroscopy. Structural characterization of devices was carried out with SEM (Helios G4 FIB-SEM) at a tilt angle of both 0° and 40°.

**Acknowledgements**

J.V.P. was supported by the National Defense Science and Engineering Graduate (NDSEG) Fellowship Program, sponsored by the Air Force Research Laboratory (AFRL), the Office of Naval Research (ONR), and the Army Research Office (ARO). J. J. C. acknowledges support from the National Science Foundation (CAREER #1749742). Device fabrication and characterization was carried out at the Yale West Campus Materials Characterization Core and the Yale West Campus Cleanroom. D.Y.Q. and A.K. acknowledge support from the National

Science Foundation under grant no. DMR 2114081. The calculations used resources of National Energy Research Scientific Computing Center (NERSC), a DOE Office of Science User Facility supported by the Office of Science of the U.S. Department of Energy under contract no. DE-AC02-05CH11231 and the Extreme Science and Engineering Discovery Environment (XSEDE), which is supported by the Office of Science of the U.S. Department of Energy under contract no. DE-AC05-00OR22725. We also thank the Yale Center for Research Computing, specifically the Grace cluster for the computing resources.

**Author contributions**

J.V.P. and J.J.C. conceived the project. J.V.P. carried out the experiments with assistance from M.W. *Ab initio* calculations were performed by A.K. and D.Y.Q. S.Y. and J.M.W. contributed to the development of experimental methods and characterization techniques. J.V.P. and J.J.C. wrote the manuscript with input from all authors.

**Competing interests**

The authors declare no competing interests.

**Supporting information**

Supplementary information is available for this manuscript.

**Figures**

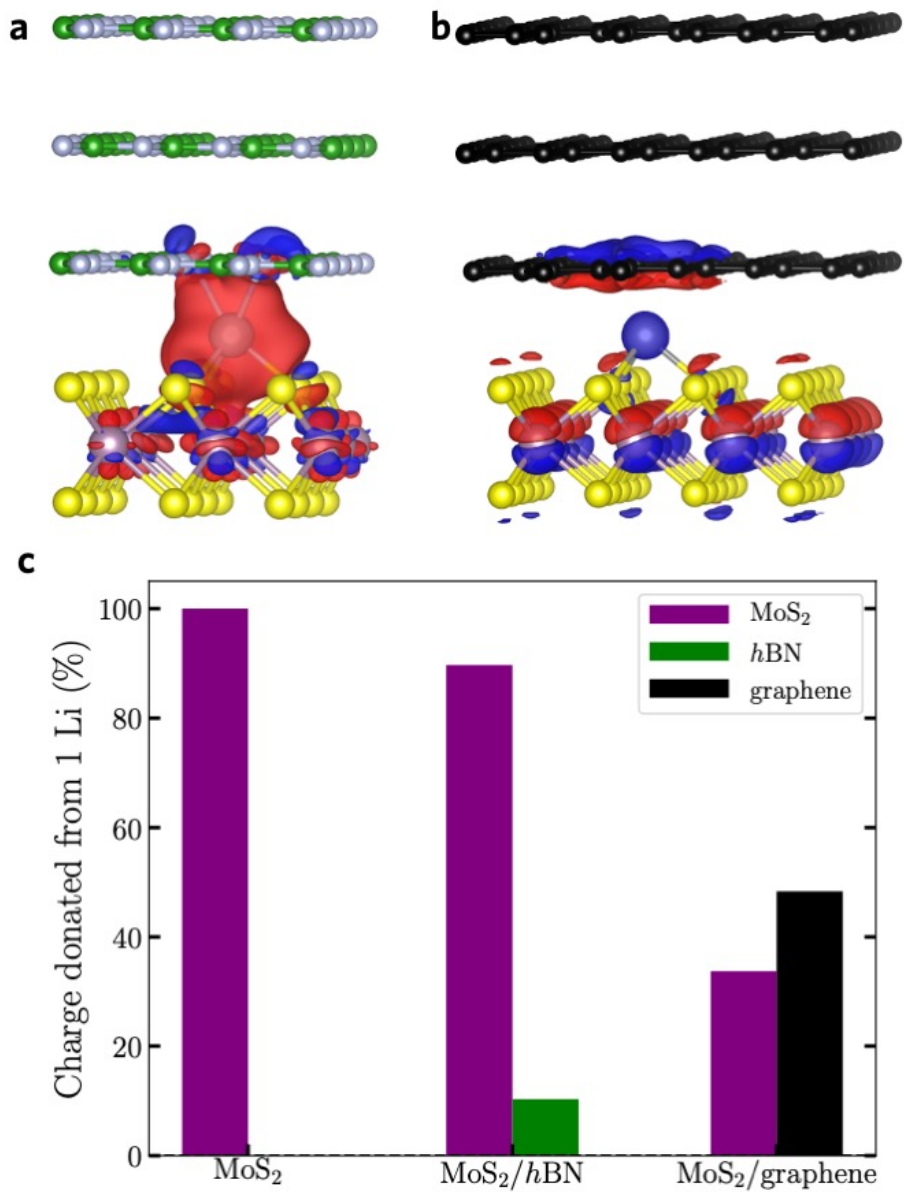

**Fig. 1 | Bader charge analysis of MoS₂ heterostructures. a-b**, Change in the charge density shown by isosurfaces containing 9% of the maximum charge density with the positive isosurfaces shown in red and the negative isosurfaces shown in blue when Li is introduced in the gap between *hBN* and MoS₂ (**a**), and the gap between graphene and MoS₂ (**b**). **c,** Bader charge analysis showing the percentage of charge donated by the Li to MoS₂ (purple) as well as *h*BN (green) and graphene (black) in the case of freestanding MoS₂ (MoS₂), the *h*BN-MoS₂ heterointerface (MoS₂/*h*BN), and the graphene-MoS₂ heterointerface (MoS₂/graphene).

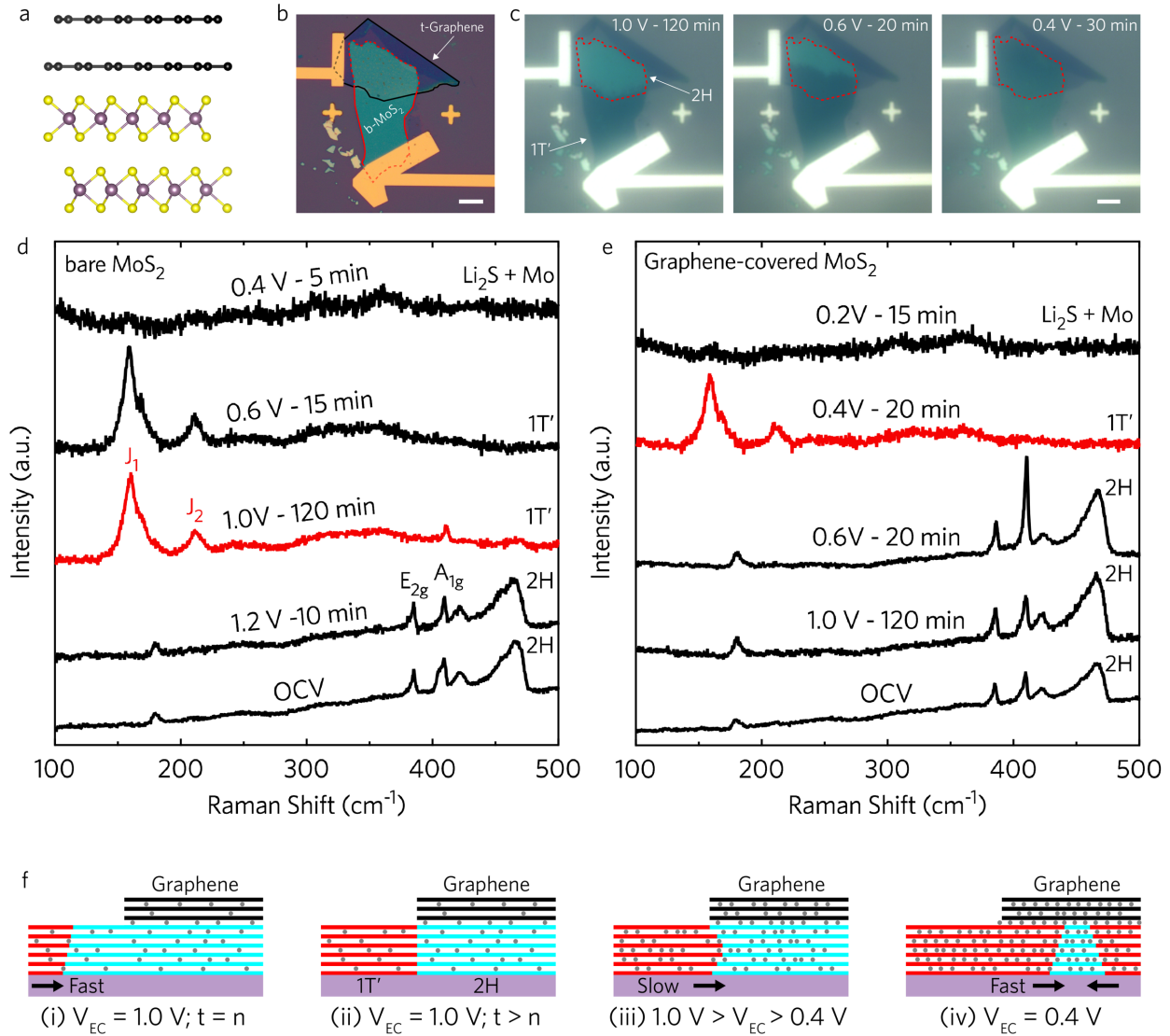

**Fig. 2 | Heterointerface-controlled nucleation of the 1T' phase in lithiated MoS$_2$. a**, Schematic cross-section of the atomic structure of the graphene-MoS$_2$ heterointerface. Mo, S, and C atoms are colored purple, yellow, and black, respectively. **b**, Optical micrograph of a MoS$_2$ flake (red outline) partially covered by a graphene flake (black outline) with gold electrical contacts; scale bar, 10 μm. **c**, *In situ* optical micrographs of the heterostructure in (**b**) during lithium intercalation at various potentials; scale bar, 10 μm. The dashed-red line outlines graphene-covered MoS$_2$. The dark color of the bare MoS$_2$ at 1.0 V vs. Li/Li$^+$ is attributed to the 1T' phase, while the graphene-covered region does not completely darken until 0.4 V vs. Li/Li$^+$. **d**, *In situ* Raman spectra from the center of the bare region of the MoS$_2$ flake in the heterostructure in (**b**) taken during intercalation. The 2H-1T' phase transition (red) is observed at 1.0 V vs. Li/Li$^+$, while the conversion reaction is observed at 0.4 V vs. Li/Li$^+$. **e**, *In situ* Raman spectra from the center of the graphene-covered region of the MoS$_2$ flake in the heterostructure in (**b**) taken during intercalation. The 2H-1T' phase transition (red) is observed at 0.4 V vs. Li/Li$^+$, while the conversion reaction is observed at 0.2 V vs. Li/Li. **f**, Schematic cross-section of the heterogenous nucleation of 1T'-MoS$_2$ (red) out of 2H-MoS$_2$ (blue) with arrows indicating the growth direction of the 1T' phase as a function of time (i-ii) and V$_{EC}$ (iii-iv).

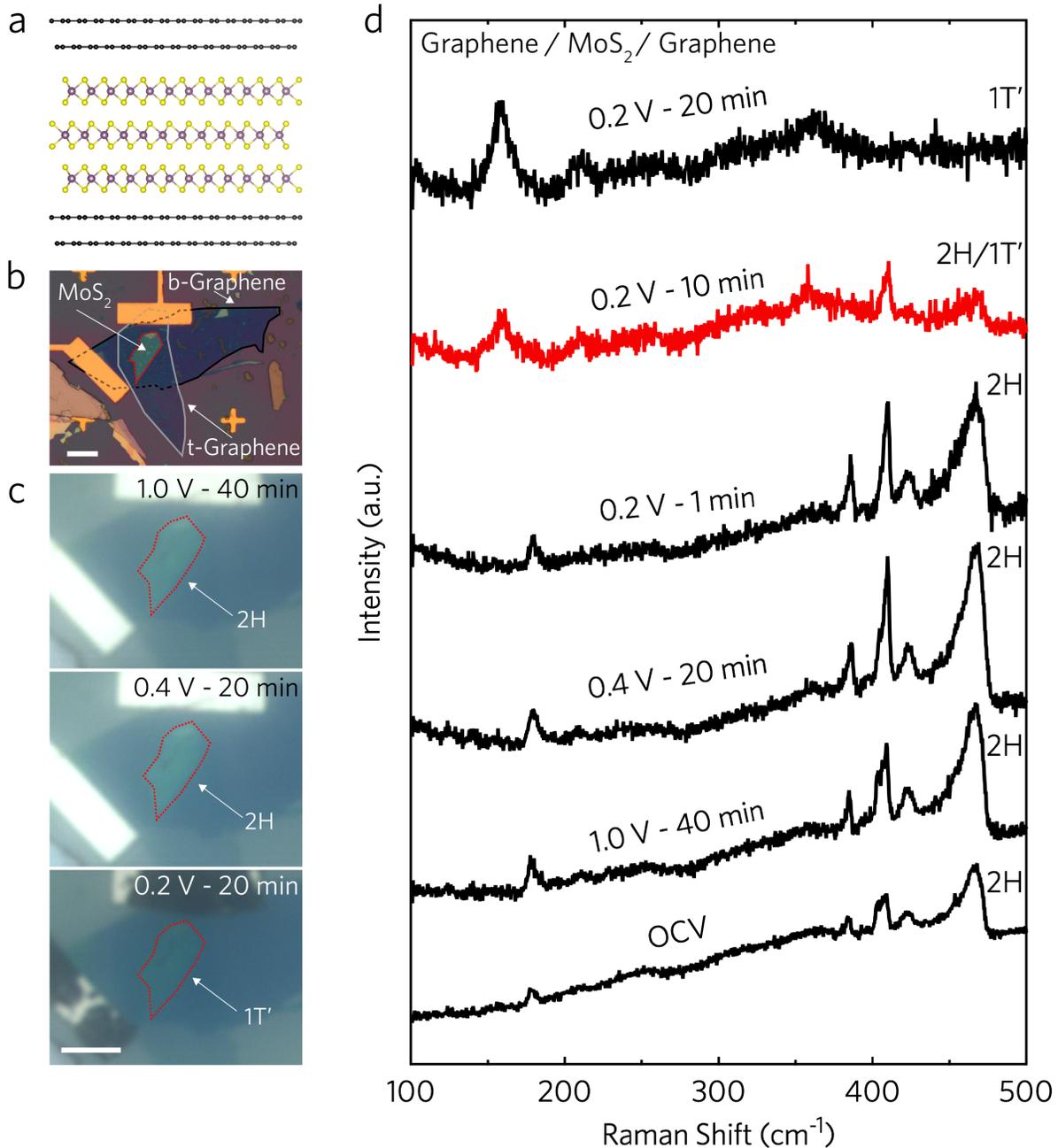

**Fig. 3 | Graphene-encapsulation of MoS₂ significantly delays the 2H-1T' phase transition. a**, Schematic cross-section of the atomic structure of a graphene/MoS₂/graphene heterostructure; Mo, S, and C atoms are colored purple, yellow, and black, respectively. In actual devices, the flakes are several layers thick. **b**, Optical micrograph of a MoS₂ flake (red outline) encapsulated by top (gray outline) and bottom (black outline) graphene flakes with gold electrical contacts; scale bar 10 μm. **c**, *In situ* optical micrographs of the heterostructure in (**b**) during lithium intercalation; scale bar, 10 μm. The dashed-red line indicates the location of the encapsulated MoS₂ flake. The MoS₂ remains unchanged until 0.2 V vs. Li/Li⁺, at which point it darkens, indicating the onset of the 1T' phase; scale bar 10 μm. The black discoloration of the gold contacts at 0.2 V vs. Li/Li⁺ indicates the alloying of gold with lithium. **d**, *In situ* Raman spectra from the center of the MoS₂ flake in the heterostructure in (**b**) taken during intercalation. The flake remains in the 2H phase down to 0.4 V vs. Li/Li⁺, and the onset of the 2H-1T' phase transition is observed only after intercalating at 0.2 V vs. Li/Li⁺ for 10 minutes, with the full transition observed after 20 minutes.

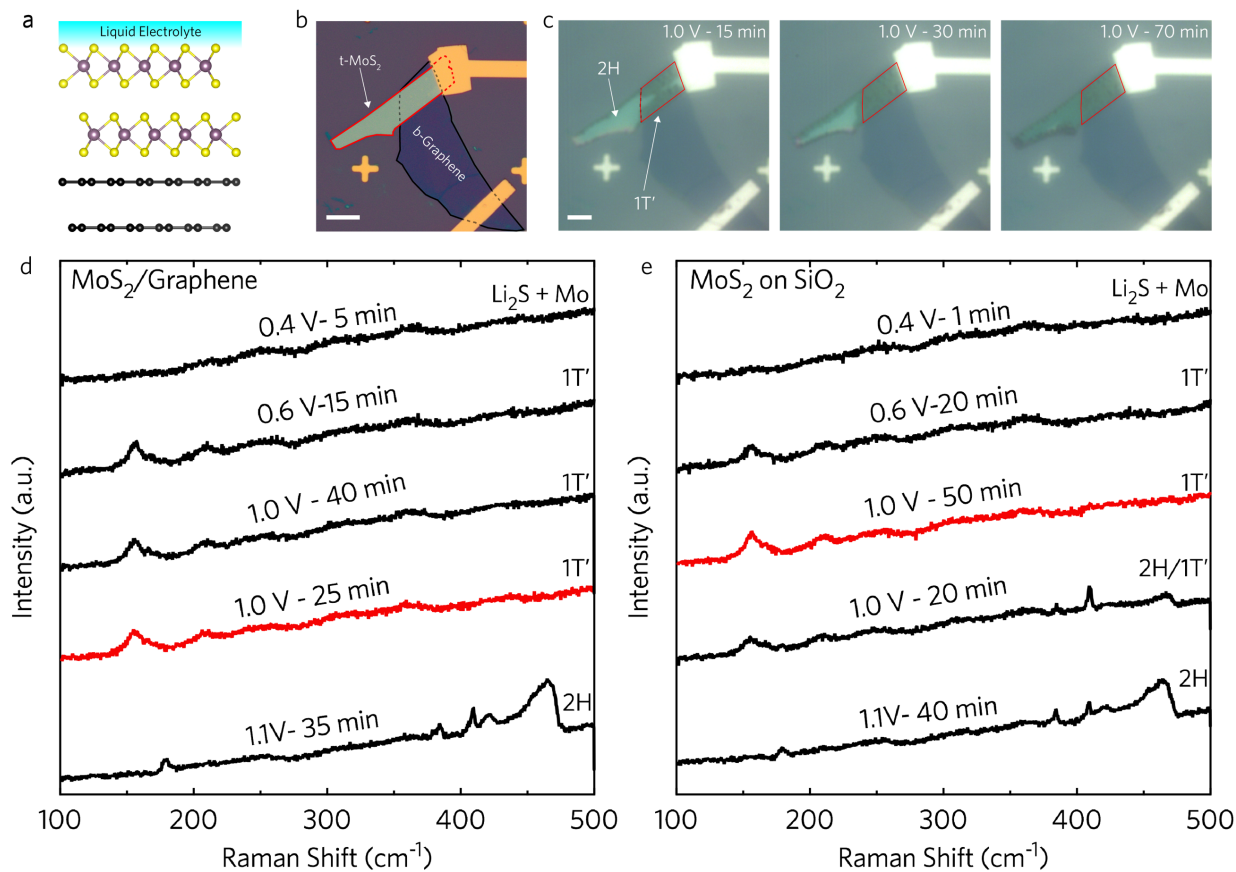

**Fig. 4 | Kinetics of the 2H-1T' phase transition in $MoS_2$. a**, Schematic cross-section of the atomic structure of a $MoS_2$/graphene heterointerface where the top basal plane of $MoS_2$ is in direct contact with liquid electrolyte; Mo, S, and C atoms are colored purple, yellow, and black, respectively. In actual devices, the flakes are several layers thick. **b**, Optical micrograph of a $MoS_2$ flake (red outline) partially supported on a graphene flake (black outline) with gold electrical contacts; scale bar, 10 μm. **c**, *In situ* optical micrographs of the heterostructure in (**b**) during lithium intercalation at various potentials; scale bar, 10 μm. The red line outlines the region of the $MoS_2$ flake supported on graphene. The dark color of $MoS_2$ at 1.0 V vs. Li/Li$^+$ is attributed to the 1T' phase, and the graphene-supported region of $MoS_2$ completes the phase transition faster than the $SiO_2$-supported region of $MoS_2$. **d**, *In situ* Raman spectra from the center of the graphene-supported region of the $MoS_2$ flake in the heterostructure in (**b**) taken during intercalation. The 2H-1T' phase transition (red) is observed after 25 minutes at 1.0 V vs. Li/Li$^+$, while the conversion reaction is observed at 0.4 V vs. Li/Li. **e**, *In situ* Raman spectra from the center of the $SiO_2$-supported region of the $MoS_2$ flake in the heterostructure in (**b**) taken during intercalation. The 2H-1T' phase transition (red) is observed after 50 minutes at 1.0 V vs. Li/Li$^+$, while the conversion reaction is observed at 0.4 V vs. Li/Li$^+$.

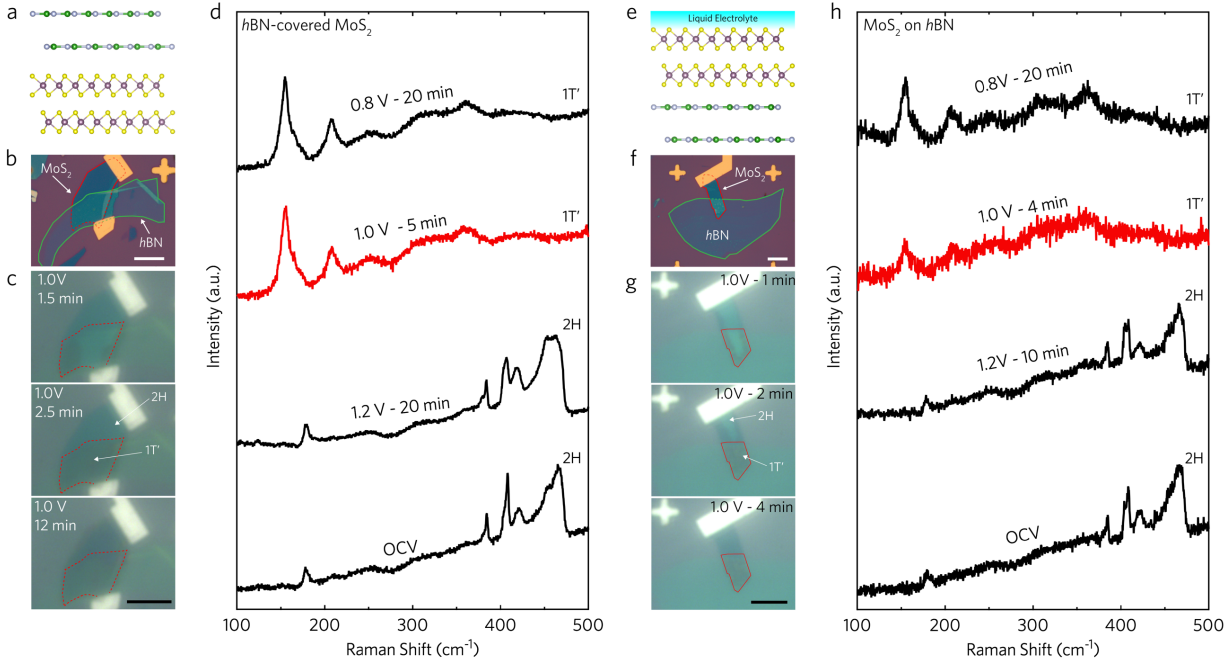

**Fig. 5 | Phase dynamics in lithiated MoS₂/hBN heterostructures. a**, Schematic cross-section of the atomic structure of the *h*BN/MoS₂ heterointerface with Mo, S, N, and B atoms colored purple, yellow, white, and green, respectively. **b**, Optical micrograph of MoS₂ (red) partially covered by *h*BN (green) with gold contacts; scale bar, 10 μm. **c**, *In situ* optical micrographs of the heterostructure in (**b**) during lithium intercalation; scale bar, 10 μm. The dashed-red line outlines *h*BN-covered MoS₂. The darkening of MoS₂ is attributed to the 1T' phase. **d**, *In situ* Raman spectra from the center of the *h*BN-covered region of MoS₂ in the heterostructure in (**b**) during intercalation. The 1T' phase (red) appears at 1.0 V vs. Li/Li⁺. **e**, Schematic cross-section of the atomic structure of a MoS₂/*h*BN heterointerface where the top basal plane of MoS₂ is in direct contact with liquid electrolyte; Mo, S, N, and B atoms are colored purple, yellow, white, and green, respectively. In actual devices, the flakes are several-layer thick. **f**, Optical micrograph of MoS₂ (red) partially supported on *h*BN (green) with gold contacts; scale bar, 10 μm. **g**, *In situ* optical micrographs of the heterostructure in (**f**) during lithium intercalation; scale bar, 10 μm. The red line outlines *h*BN-supported MoS₂. The darkening of MoS₂ is attributed to the 1T' phase. **h**, *In situ* Raman spectra from the center of the *h*BN-supported region of MoS₂ in the heterostructure in (**f**) during intercalation. The 1T' phase (red) appears at 1.0 V vs. Li/Li⁺.